\documentclass[A4]{emulateapj} \usepackage{apjfonts}

\shorttitle{Black holes in galaxy mergers: Formation
of red elliptical galaxies}

\shortauthors{Springel, Di~Matteo, \& Hernquist}

\slugcomment{Submitted to the Astrophysical Journal Letters}

\begin{document}        

\title{Black holes in galaxy mergers: The
  formation of red elliptical galaxies}

\author{Volker~Springel\altaffilmark{1}, Tiziana~Di~Matteo\altaffilmark{1}, 
and Lars~Hernquist\altaffilmark{2}}

\altaffiltext{1}{Max-Planck-Institut f\"ur Astrophysik, Garching, Germany}
\altaffiltext{2}{Harvard-Smithsonian Center for Astrophysics, Cambridge, USA}

\begin{abstract}
  We use hydrodynamical simulations to study the color transformations
  induced by star formation and active galactic nuclei (AGN) during
  major mergers of spiral galaxies.  Our modeling accounts for
  radiative cooling, star formation, and supernova feedback.
  Moreover, we include a treatment of accretion onto supermassive
  black holes embedded in the nuclei of the merging galaxies. We
  assume that a small fraction of the bolometric luminosity of an
  accreting black hole couples thermally to surrounding gas, providing
  a feedback mechanism that regulates its growth. The encounter and
  coalescence of the galaxies triggers nuclear gas inflow which fuels
  both a powerful starburst and strong black hole accretion.
  Comparing simulations with and without black holes, we show that AGN
  feedback can quench star formation and accretion on a short
  timescale, particularly in large galaxies where the black holes can
  drive powerful winds once they become sufficiently massive.  The
  color evolution of the remnant differs markedly between mergers with
  and without central black holes.  Without AGN, gas-rich mergers lead
  to ellipticals which remain blue owing to residual star formation,
  even after more than 7 Gyrs have elapsed. In contrast, mergers with
  black holes produce ellipticals that redden much faster, an effect
  that is more pronounced in massive remnants where a nearly complete
  termination of star formation occurs, allowing them to redden to
  $u-r\simeq 2.3$ in less than one Gyr.  AGN feedback may thus be
  required to explain the population of extremely red massive early
  type-galaxies, and it appears to be an important driver in
  generating the observed bimodal color distribution of galaxies in
  the Local Universe.
\end{abstract}


\section{Introduction}

In hierarchical theories of galaxy formation, large systems are built
up from mergers of smaller progenitors.  Direct support for this
picture comes from interacting pairs of galaxies seen in the Local
Universe \citep{Toomre72}.  Fully self-consistent numerical models
have demonstrated that interactions and mergers of spiral galaxies can
produce remnants with properties similar to large elliptical galaxies
\citep[e.g.][]{Barnes88,Barnes92,Hernquist92,Hernquist93}, as expected
according to the ``merger hypothesis'' \citep{Toomre77}.

However, it is still controversial whether the merger scenario can
account for detailed properties of the local galaxy population.  For
example, from large surveys like SDSS, 2dFGRS or DEEP, it has been
shown that the color distribution at fixed luminosity is bimodal
\citep[e.g.][]{Strateva2001,Blanton2003,Kauffmann2003a}, and can be well
fitted by two Gaussians \citep[e.g.][]{Baldry2004}. The mean and
variance of these two distributions depend on luminosity, but
surprisingly little on galaxy environment \citep{Balogh2004}. Also,
there exists a population of massive, very red galaxies even at high
redshift \citep[e.g.][]{Franx2003}, which has been interpreted as
evidence for monolithic galaxy formation at early times, rather than a
more gradual build-up by a sequence of mergers.

If mergers of galaxies indeed produce red ellipticals from blue,
star-forming disks, the color must be transformed from red to blue on
a relatively short timescale, otherwise the `gap' between the blue and
red distributions would be washed out. The most straightforward way to
achieve a rapid reddening of an elliptical would be for star formation
to terminate abruptly following a merger, as could be the case, for
example, if all the gas were consumed in a starburst.

It is, however, unclear whether merger-induced starbursts necessarily
consume all the available gas, particularly in gas-rich mergers at
high redshift.  If they fail to do so and a small fraction of the gas
remains, even a relatively low level of star formation in the remnant
will prevent it from reaching the extremely red colors characteristic
of many ellipticals. Instead, the residual star formation would
decline slowly over a Hubble time \citep[e.g.][]{MH94, MH96, HM95},
and the remnant would make a gradual transition into the red
population that would blur the observational distinction between red
and blue galaxies.

Here, we use hydrodynamical simulations of gas-rich mergers without
AGN feedback to show that they do not necessarily produce remnants
that are extremely gas-poor, even if a powerful starburst consumes a
substantial fraction of the gas. Consequently, the color of the
remnants does not evolve sufficiently rapidly to be consistent with
the mean of the red population of the observed bimodal color
distribution.

The situation is very different when the impact of central AGN in the
merging galaxies is included. In recent years, a remarkable connection
between galaxy formation and supermassive black holes has been
revealed, indicating that their growth is linked.  Perhaps the most
direct evidence for this is the correlation seen between the stellar
velocity dispersion of bulges and the masses of the black holes they
host \citep[e.g.][]{Tremaine2002}.  Theoretical models conjecture that
the correlation arises because black hole growth stalls once the
energy deposition associated with the accretion can expel the
remaining gas from the halo or bulge
\citep{Ciotti97,Ciotti2001,Silk98,Wyithe2003}.  This would also have
an immediate bearing on star formation in the galaxy.  In our
simulations we include black hole accretion and feedback to examine
their impact on star formation during galaxy mergers, focusing on the
color evolution of the ensuing ellipticals.  Our results demonstrate
that these processes can quench star formation in large merger
remnants, and that ellipticals formed in this manner redden
sufficiently rapidly to explain the observed color bimodality of local
galaxies.

\section{Numerical Simulations}

Our simulations were performed with {\small GADGET-2}, a new version
of the parallel TreeSPH code {\small GADGET} \citep{Springel2001}.  It
uses an entropy-conserving formulation of SPH \citep{Springel2002},
and includes radiative cooling, heating by a UV background, and a
sub-resolution model of the multiphase structure of dense gas to
describe star formation and supernova feedback \citep{Springel2003}.

We have incorporated a novel procedure for handling accretion onto
supermassive black holes (BHs) into this code.  Briefly, we represent
BHs by ``sink'' particles that accrete gas from their local
environment. The accretion rate $\dot M_{\rm B}$ is estimated from the
local gas density and sound speed, using a Bondi-Hoyle-Lyttleton
parameterization together with an imposed upper bound equal to the
Eddington rate. We further assume that a small fraction of 5\% of the
bolometric luminosity of $0.1 \dot M_{\rm B} c^2$ (for an accretion
efficiency of 10\%) can couple dynamically to the ambient gas around
the accreting black hole. This source of feedback is injected as
thermal energy into the gas around the BH particle.  A full discussion
of our methodology is given in Springel, Di Matteo \& Hernquist
(2004). See also the recent studies by \citet{Kawata2004}, who
investigated a thermal AGN heating model not coupled directly to 
accretion, and by \citet{Kazantzidis2004}, who followed completely
`passive' black hole particles in galaxy mergers.

\begin{figure}[t]
\begin{center}
\resizebox{8.0cm}{!}{\includegraphics{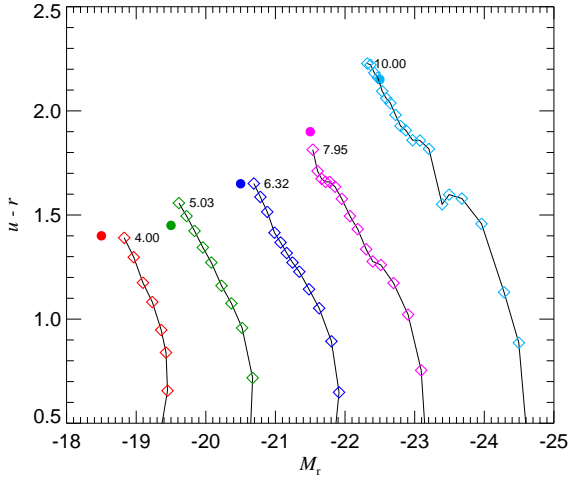}}\vspace*{-0.4cm}\\
\end{center}
\caption{Evolutionary tracks of isolated, star-forming spiral galaxies
 initially with pure gas disks, in the color-magnitude plane of $u-r$
 vs.~$M_r$. From left to right, are models with $V_{\rm vir}= 80,$
 113, 160, 226, and $320\,{\rm km\, s^{-1}}$.  Diamonds on each track
 are spaced 0.5 Gyrs apart, and the age of the last point is labeled.
 Filled circles show the mean color of the blue part of the observed
 bimodal color distribution at a given luminosity \citep{Balogh2004}.
\label{FigIsolated}}
\end{figure}

Note that we do not attempt to resolve the small-scale accretion
dynamics near the black hole; i.e.~the complex processes that are
ultimately responsible for transporting gas down to the last stable
orbit. Instead, our modeling is based on the assumption that the
time-averaged accretion, and feedback associated with the accretion,
can be estimated from properties of the gas on scales $\sim 100\,{\rm
pc}$, similar to our spatial resolution.

We generate stable, isolated disk galaxies using the approach outlined
in \citet{Springel2004}. Each galaxy has an extended dark matter halo
with a profile motivated by cosmological simulations, an exponential
disk of gas and stars, and a bulge.  Here, we focus on only one
particular choice for the structural properties of our disk galaxies,
noting that our results are relatively insensitive to the details of
these choices. We do, however, consider various galaxy masses,
yielding a family of self-similar disk galaxy models with virial
velocities $V_{\rm vir}=80$, $113$, $160$, $226$, and $320\,{\rm
km\,s^{-1}}$. The total mass of each galaxy is $M_{\rm vir}= V_{\rm
vir}^3/(10 G H_0)$, with the baryonic disk having a mass fraction of
$m_{\rm d}=0.041$, the bulge of $m_{\rm b}= 0.0136$, and the rest
being dark matter. The scale length of the disk is computed based on
an assumed spin parameter of $\lambda=0.041$, and the scale-length of
the bulge set to 0.2 times the resulting disk scale-length.  For a
fiducial choice of $V_{\rm vir}=160\,{\rm km\,s^{-1}}$, the rotation
curve and mass of the resulting model galaxy is similar to the Milky
Way. Note that without the scale-dependent physics of cooling, star
formation and black hole accretion, these galaxies would evolve in a
self-similar fashion.

We eliminate the initial gas fraction of the disks as a separate free
parameter by starting our simulations with pure gaseous disks. We here
take advantage of the ability of our sub-resolution model for the
star-forming gas to stably evolve even massive gaseous disks.  This
also avoids the need to specify an age distribution for disk stars
that may already be present initially.  In our default models, we use
168000 particles for the dark matter halo, 8000 particles for the
bulge, 24000 particles for the gaseous disk, and one black hole sink
particle, if present. The latter is given an initial seed mass of
$10^5\,{\rm M}_\odot$ in all simulations. With this choice, the dark
matter particles, gas particles, and star particles are all roughly of
equal mass, and the central cusps in the profiles for the dark matter
and bulge \citep{hern90} are reasonably well resolved. To check
numerical convergence, we have also run a few of our simulations with
eight times as many particles, where each galaxy model has 1.5 million
particles in total.

We have carried out two different types of simulations. In our first
set of runs, we evolved the galaxies in isolation to study how the gas
disks are turned into stars, providing a simple model for the color
evolution of quiescent, star-forming disk galaxies. In a second set,
we have used pairs of the same models and set them on a collision
course, with zero orbital energy and a small pericenter separation of
$7.1\,{\rm kpc}$ (for the $V_{\rm vir}=160\,{\rm km\,s^{-1}}$
case). By varying the initial separation in some of our mergers, we
have also changed the time until the first encounter of the galaxies,
and hence effectively modified the gas fraction in the galaxies when
they coalesce. In the simulations analyzed here we only consider pure
prograde encounters, for simplicity. However, we have checked with
further simulations of more general encounters, where the disk spin
vectors were tilted relative to the orbital plane, that our results
are insensitive to the orbital configuration.

\begin{figure}[t]
\begin{center}
\resizebox{8.0cm}{!}{\includegraphics{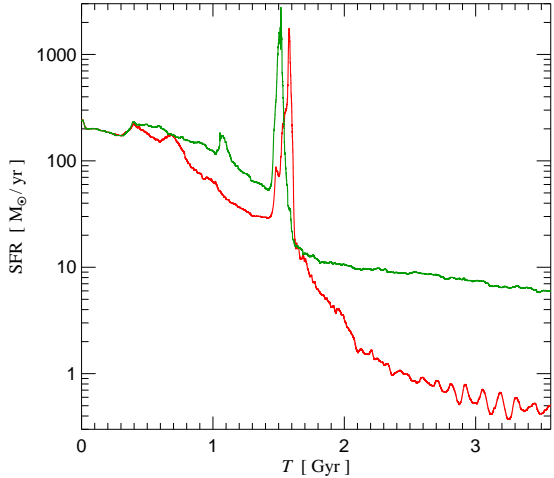}}\vspace*{-0.4cm}\\
\end{center}
\caption{Comparison of the star formation rate history of two
colliding gas-rich spirals of mass $3.85\times 10^{12}\,{\rm M}_\odot$
with (red line) and without (green line) central supermassive BHs. The
merger triggers a powerful starburst at time $\sim 1.5\,{\rm Gyr}$,
which is accompanied by a phase of Eddington accretion in the
simulation with BHs. The feedback energy from accretion eventually
blows away the gas surrounding the black holes, nearly terminating
star formation in the remnant and stalling further growth of the black
holes.
\label{FigCompSfr}}
\end{figure}

\section{Results}

\subsection{Color evolution of isolated disk galaxies}

In Figure~\ref{FigIsolated}, we show evolutionary tracks of isolated,
star-forming spiral galaxies in the color-magnitude plane of $u-r$
vs.~$M_r$.  We use the stellar population synthesis models of
\citet{Bruzual2003} to compute rest-frame magnitudes in the SDSS bands
for the simulated galaxies, assuming solar metallicity and a Chabrier
initial mass function. We do not add corrections for internal
extinction in the galaxies.

The galaxies start with pure gaseous disks, which are transformed into
stellar disks on roughly an exponential timescale.  The more massive
galaxies shown in Figure~\ref{FigIsolated} have somewhat shorter gas
consumption timescales than less massive ones, as expected from the
density-dependence of the assumed Schmidt-like star formation law
\citep{Springel2003}. This makes them slightly redder at the same age.
However, to reproduce the strong trend with luminosity seen in the
mean color of the blue population of star-forming galaxies
\citep{Balogh2004}, one needs to assume that larger galaxies are also
older. Formally, we obtain a good match to the observed trend if
galaxies of total mass $\sim 4\times 10^{12}\,{\rm M}_\odot$ started
forming their stars about $4\,{\rm Gyr}$ earlier than galaxies of mass
$\sim 10^{11}\,{\rm M}_\odot$. While this trend qualitatively agrees
with the proposed notion of `cosmic down-sizing' of star formation
\citep{Cowie1996,Kauffmann2003b}, it is important to note that our isolated
systems represent at best a crude model for disk formation because
several cosmological effects are neglected, most notably infall. The
results therefore primarily serve to illustrate the color evolution of
our galaxies when they do not suffer a merger.

\subsection{Star formation and color evolution in mergers}

In Figure~\ref{FigCompSfr}, we compare the star formation rates in
collisions between two large gas-rich spirals, with and without black
holes.  The collision causes a nuclear inflow of gas, triggering a
strong starburst, and fueling black hole accretion in the simulation
with AGN. The feedback resulting from accretion first only damps the
starburst, but once the black hole has accreted at its Eddington rate
for several Salpeter times, it begins to drive a powerful quasar
outflow. This wind can remove much of the gas from the inner regions
of the merging galaxies, thereby nearly terminating star formation on
a short timescale. As a result, there is almost no residual star
formation in the remnant with black holes, as opposed to the ordinary
simulation where the remnant keeps forming stars at a non-negligible
rate of a few ${\rm M_\odot/yr}$ for several Gyr.  An analysis of the
dynamics and the final masses of the BHs in these simulations is given
in \citet{DiMatteo2004}.

\begin{figure}[t]
\begin{center}
\resizebox{8cm}{!}{\includegraphics{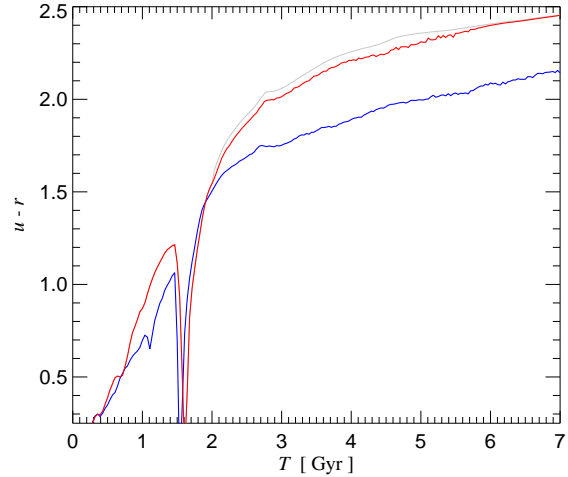}}\vspace*{-0.4cm}\\%
\end{center}
\caption{Comparison of the color evolution of the merger of two
colliding gas-rich spirals of mass $3.85\times 10^{12}\,{\rm M}_\odot$
($V_{\rm vir}=226\,{\rm km\,s}^{-1}$) with (red line) and without
(blue line) central supermassive BHs.  The thin gray line marks a
fiducial color evolution assuming that no stars are formed after
$T=2\,{\rm Gyr}$.
\label{FigEvolMergers}}
\end{figure}

In Figure~\ref{FigEvolMergers}, we compare the temporal evolution of
the $u-r$ color in these two merger simulations. After a brief
excursion into the extreme blue during the bursts, when much of the
gas is consumed, both remnants begin to redden. However, this happens
substantially faster when AGN feedback is included.  In fact, in our
simulation set we find that for galaxies more massive than $\sim
3\times 10^{12}\,{\rm M}_\odot$ the color evolution of the remnants is
consistent with one where no stars are formed {\em at all} after the
burst -- they will hence quickly evolve into extremely red, massive
elliptical galaxies.

However, we note that the magnitude of this ``termination effect''
depends on the masses of the galaxies involved.  Because the BHs in
small galaxies grow only relatively little in mass, consistent with
the $M_{\rm B}-\sigma$ relation, AGN feedback is much less efficient
in smaller galaxies.  Consequently, the change in the remnant
evolution is progressively weaker for less massive galaxies.  In the
smallest galaxies we considered, of virial velocity $V_{\rm
vir}=80\,{\rm km\,s^{-1}}$, the color evolution is nearly unchanged
between simulations with and without black holes. In mergers of these
systems, the small spheroidal galaxies that form remain relatively
gas-rich and exhibit ongoing star formation. Such galaxies appear to
exist. For example, using data from the DEEP survey of the Groth
strip, \citet{Im2001} show that a substantial fraction of
morphologically selected early-type galaxies at $z\le 1$ have blue
colors, and that they are likely to be low-mass, star-forming
spheroids.

\subsection{Relation to the bimodal color distribution}

\begin{figure}[t]
\begin{center}
\resizebox{8.0cm}{!}{\includegraphics{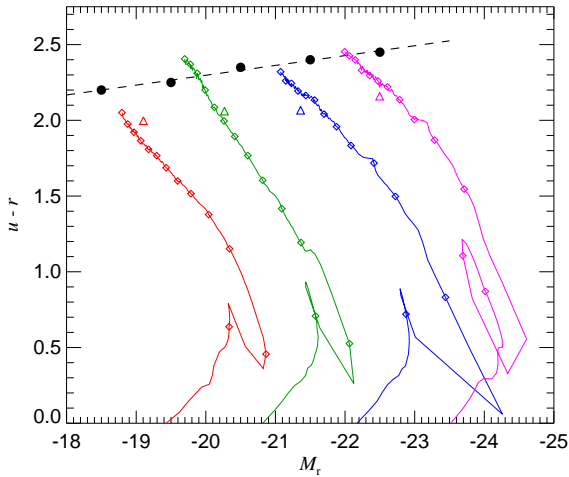}}\vspace*{-0.4cm}\\
\end{center}
\caption{Color evolution in the $u-r$ vs. $M_r$ plane for gas-rich
mergers with black hole accretion. Symbols on the tracks are spaced
0.5 Gyr apart, with the last point corresponding to an age of
$\sim~5.5\,{\rm Gyr}$ after the merger-induced starburst.  For
comparison, triangles show mergers without black holes at the same
time, and the solid circles give the observed mean color of the red
part of the bimodal color distribution at a given luminosity
\citep{Balogh2004}.}
\label{FigMergerColorEvol}
\end{figure}

In Figure~\ref{FigMergerColorEvol}, we show evolutionary tracks of the
color evolution of the merger simulations for different progenitor
masses, again in the $u-r$ vs. $M_r$ plane.  The last points on the
tracks correspond to an age of $\sim~5.5\,{\rm Gyr}$ after the
merger-induced starbursts. At this fiducial time, we compare to the
mean color of the red part of the bimodal color distribution in the
Local Universe, as determined by \citet{Balogh2004} for the SDSS.

The large spacings of the markers on the track of the massive disk
galaxies during the transition from blue to red illustrate how rapidly
the color transformation proceeds. Already $\sim 1\,{\rm Gyr}$ after
the merger, the color has reddened to about $u-r\simeq 2.0$, and after
a further Gyr, it reaches about $u-r\simeq 2.2$. In contrast, without
black holes the remnant takes $5.5\,{\rm Gyr}$ to reach $u-r\simeq
2.1$, and has difficulty reaching the observed redness even after a
Hubble time.

This result also demonstrates an important connection to the observed
bimodal color distribution of galaxies. AGN feedback appears to be
required in order to move galaxies from the blue star-forming
population into the red population of ``dead'' galaxies sufficiently
rapidly. If the transition is too slow, there should be many more
galaxies with intermediate colors, which would wash out the observed
bimodality.  Interestingly, the observed trend with luminosity of the
mean color of the red mode of the bimodal color distribution can be
approximately reproduced by our merger remnants with BHs, at a time
roughly $5.5\,{\rm Gyr}$ after completion of the mergers. As a
look-back time, this would correspond to a formation redshift of
$z\simeq 0.7$. Without black holes, the galaxies reach the required
redness only much later, or not at all within a Hubble time.  While
the idealized nature of our individual galaxy mergers preclude us from
drawing definite statistical conclusions, our results indicate that BH
feedback is essential for shaping the bimodal color distribution of
galaxies.

\section{Conclusions}

We have demonstrated that gas-rich galaxies do not necessarily consume
all their gas in the starbursts that accompany major mergers.
Consequently, ellipticals formed in such events can sustain star
formation for extended periods of many Gyrs that makes them relatively
blue.  However, if the merging galaxies host supermassive black holes
at their centers, AGN feedback provides a mechanism to quench star
formation on a short timescale. This introduces a marked difference in
the color evolution of galaxies: mergers of massive galaxies can
produce remnants that redden to $u-r\simeq 2.2-2.3$ in about $1-2\,
{\rm Gyrs}$. Moreover, the AGN feedback drives a gaseous outflow which
leaves behind a gas-poor remnant. The ``dead'' ellipticals formed in
this manner should be a good match to the luminous red stellar
populations of many massive ellipticals, which are devoid of
star-forming gas and lack young stars. Also, AGN feedback may be an
important driver in shaping the observed bimodal color distribution of
galaxies.

Because black hole growth is a strong function of the size of the
spheroid formed, the effects of AGN feedback sensitively depend on the
masses of the merging galaxies. In our simulations, black hole
accretion modifies the properties of large elliptical remnants
strongly, while those of forming dwarf spheroidal systems are largely
unaffected.

It is now widely believed that the formation of spheroids and the
growth of supermassive black holes are intimately linked.  If this is
the case, then black holes should not be ignored in models of galaxy
formation, since even basic properties like the color of ellipticals
can be influenced strongly by them, as we have shown here.
Hydrodynamical simulations of galaxy formation that self-consistently
account for star formation and the growth of black holes promise to be
an important tool for exploring this connection further, which may
well lead to fundamental changes in the theory of hierarchical galaxy
formation.

\vspace*{-0.2cm}
\acknowledgments This work was supported in part by NSF grants ACI
96-19019, AST 00-71019, AST 02-06299, and AST 03-07690, and NASA ATP
grants NAG5-12140, NAG5-13292, and NAG5-13381.  The simulations were
performed at the Center for Parallel Astrophysical Computing at the
Harvard-Smithsonian Center for Astrophysics.

\bibliographystyle{apj}
\bibliography{springel}

\begin{thebibliography}{19}
\expandafter\ifx\csname natexlab\endcsname\relax\def\natexlab#1{#1}\fi

\bibitem[{{Baldry} {et~al.}(2004){Baldry}, {Glazebrook}, {Brinkmann}, {Ivezi{\'
  c}}, {Lupton}, {Nichol}, \& {Szalay}}]{Baldry2004}
{Baldry}, I.~K., {Glazebrook}, K., {Brinkmann}, J., {Ivezi{\' c}}, {\v Z}.,
  {Lupton}, R.~H., {Nichol}, R.~C., \& {Szalay}, A.~S. 2004, \apj, 600, 681

\bibitem[{{Balogh} {et~al.}(2004){Balogh}, {Baldry}, {Nichol}, {Miller},
  {Bower}, \& {Glazebrook}}]{Balogh2004}
{Balogh}, M.~L., {Baldry}, I.~K., {Nichol}, R.~C., {Miller}, C., {Bower}, R.,
  \& {Glazebrook}. 2004, \apj, in press, astro-ph/0406266

\bibitem[{{Barnes}(1988)}]{Barnes88}
{Barnes}, J.~E. 1988, \apj, 331, 699

\bibitem[{{Barnes}(1992)}]{Barnes92}
---. 1992, \apj, 393, 484

\bibitem[Blanton et~al.(2003)]{Blanton2003}
Blanton, M.~R., et al. 2003, ApJ, 594, 186

\bibitem[{{Bruzual} \& {Charlot}(2003)}]{Bruzual2003}
{Bruzual}, G. \& {Charlot}, S. 2003, \mnras, 344, 1000

\bibitem[{{Ciotti} \& {Ostriker}(1997)}]{Ciotti97}
{Ciotti}, L. \& {Ostriker}, J.~P. 1997, \apjl, 487, L105

\bibitem[{{Ciotti} \& {Ostriker}(2001)}]{Ciotti2001}
---. 2001, \apj, 551, 131

\bibitem[Cowie et al. (1996)]{Cowie1996}
Cowie, L.~L., Songaila, A., Hu, E.~M., Cohen, J.~G. 1996, AJ, 112, 839

\bibitem[{{Di Matteo} {et~al.}(2004){Di Matteo}, {Springel}, \&
  {Hernquist}}]{DiMatteo2004}
{Di Matteo}, T., {Springel}, V., \& {Hernquist}, L. 2004, in preparation

\bibitem[{{Franx} {et~al.}(2003){Franx}, {Labb{\' e}}, {Rudnick}, {van Dokkum},
  {Daddi}, {F{\" o}rster Schreiber}, {Moorwood}, {Rix}, {R{\" o}ttgering}, {van
  de Wel}, {van der Werf}, \& {van Starkenburg}}]{Franx2003}
{Franx}, M., et al. 2003, \apjl, 587, L79

\bibitem[{{Hernquist}(1990){Hernquist}}]{hern90}
{Hernquist}, L. 1990, \apj, 356, 359

\bibitem[{{Hernquist}(1992)}]{Hernquist92}
---. 1992, \apj, 400, 460

\bibitem[{{Hernquist}(1993)}]{Hernquist93}
---. 1993, \apj, 409, 548

\bibitem[Hernquist \& Mihos(1995)]{HM95}
Hernquist, L. \& Mihos, J.C. 1995, \apj, 448, 41

\bibitem[{{Im} {et~al.}(2001){Im}, {Faber}, {Gebhardt}, {Koo}, {Phillips},
  {Schiavon}, {Simard}, \& {Willmer}}]{Im2001}
{Im}, M., {Faber}, S.~M., {Gebhardt}, K., {Koo}, D.~C., {Phillips}, A.~C.,
  {Schiavon}, R.~P., {Simard}, L., \& {Willmer}, C.~N.~A. 2001, \aj, 122, 750

\bibitem[Kauffmann {et~al.}(2003a)]{Kauffmann2003a}
{Kauffmann}, G., et al. 2003a, \mnras, 241, 54


\bibitem[Kauffmann {et~al.}(2003b)]{Kauffmann2003b}
{Kauffmann}, G., et al. 2003b, \mnras, 341, 54

\bibitem[Kazantzidis et al.(2004)]{Kazantzidis2004}
Kazantzidis, S. et al. 2004, \apj, submitted, astro-ph/0407407

\bibitem[Kawata \& Gibson (2004)]{Kawata2004}
Kawata, D., Gibson, B.~K. 2004, \mnras, submitted, astro-ph/0409068

\bibitem[Mihos \& Hernquist(1994)]{MH94}
Mihos, J.C. \& Hernquist, L. 1994, \apj, 425, L13

\bibitem[Mihos \& Hernquist(1996)]{MH96}
Mihos, J.C. \& Hernquist, L. 1996, \apj, 464, 641

\bibitem[{{Silk} \& {Rees}(1998)}]{Silk98}
{Silk}, J. \& {Rees}, M.~J. 1998, \aap, 331, L1

\bibitem[{{Springel} {et~al.}(2004){Springel}, {Di Matteo}, \&
  {Hernquist}}]{Springel2004}
{Springel}, V., {Di Matteo}, T., \& {Hernquist}, L. 2004, in preparation

\bibitem[{{Springel} \& {Hernquist}(2002)}]{Springel2002}
{Springel}, V. \& {Hernquist}, L. 2002, \mnras, 333, 649

\bibitem[{{Springel} \& {Hernquist}(2003)}]{Springel2003}
---. 2003, \mnras, 339, 289

\bibitem[{{Springel} {et~al.}(2001){Springel}, {Yoshida}, \&
  {White}}]{Springel2001}
{Springel}, V., {Yoshida}, N., \& {White}, S.~D.~M. 2001, New Astronomy, 6, 79

\bibitem[Strateva et~al.(2001)]{Strateva2001}
Strateva, I., et al. 2001, AJ, 122, 1861

\bibitem[Toomre(1977)]{Toomre77}
Toomre, A. 1977, in The Evolution of Galaxies and Stellar 
Populations, ed. B.M. Tinsley \& R.B. Larson (Yale Univ. Obs: New Haven),
p. 401

\bibitem[{{Toomre} \& {Toomre}(1972)}]{Toomre72}
{Toomre}, A. \& {Toomre}, J. 1972, \apj, 178, 623

\bibitem[{{Tremaine} {et~al.}(2002){Tremaine}, {Gebhardt}, {Bender}, {Bower},
  {Dressler}, {Faber}, {Filippenko}, {Green}, {Grillmair}, {Ho}, {Kormendy},
  {Lauer}, {Magorrian}, {Pinkney}, \& {Richstone}}]{Tremaine2002}
{Tremaine}, S., et al. 2002, \apj, 574, 740

\bibitem[{{Wyithe} \& {Loeb}(2003)}]{Wyithe2003}
{Wyithe}, J.~S.~B. \& {Loeb}, A. 2003, \apj, 595, 614

\end{thebibliography}

\end{document}